# BioBlender: Fast and Efficient All Atom Morphing of Proteins Using Blender Game Engine

**Maria Francesca Zini[1], Yuri Porozov[1,2], Raluca Mihaela Andrei[1,3], Tiziana Loni[4], Claudia Caudai[1] and Monica Zoppè[1,§]**

[1]Scientific Visualization Unit, Institute of Clinical Physiology, CNR of Italy, Area della Ricerca, Pisa, Italy
[2] present address: Univ of St. Petersburg, Russia
[3]Scuola Normale Superiore, Pisa, Italy
[4]Big Bang Solutions, Navacchio (Pisa), Italy
[§]Corresponding author

Email addresses:
MFZ: myrtil@gmail.com
YP: porozov@sns.it
RMA: r.andrei@sns.it
TL: tialo@tiscali.it
CC: claudia.caudai@gmail.com
MZ: mzoppe@ifc.cnr.it

## Abstract

In this and the associated article *BioBlender: A Software for Intuitive Representation of Surface Properties of Biomolecules* [1], we present BioBlender as a complete instrument for the elaboration of motion (here) and the visualization [1] of proteins and other macromolecules, using instruments of computer graphics.
A vast number of protein (if not most) exert their function through some extent of motion. Despite recent advances in higly performant methods, it is very difficult to obtain direct information on conformational changes of molecules. However, several systems exist that can shed some light on the variability of conformations of a single peptide chain; among them, NMR methods provide collections of a number of static 'shots' of a moving protein.
Starting from this data, and assuming that if a protein exists in more than 1 conformation it must be able to transit between the different states, we have elaborated a system that makes ample use of the computational power of 3D computer graphics technology. Considering information of all (heavy) atoms, we use animation and game engine of Blender to obtain transition states.
The model we chose to elaborate our system is Calmodulin, a protein favorite among structural and dynamic studies due to its (relative) simplicity of structure and small dimension. Using Calmodulin we show a procedure that enables the building of a 'navigation map' of NMR models, that can help in the identification of movements. In the process, a number of intermediate conformations is generated, all of which respond to strict bio-physical and bio-chemical criteria. The BioBlender system is available for download from the website www.bioblender.net, together with examples, tutorial and other useful material.

## Introduction

The processes that constitute life are the result of a coordinated sequence of events involving cellular components. Among the most active, flexible and versatile of all cellular macromolecules are proteins. The long-standing paradigm of 'structure-function' has been enriched, in recent years, with the increasing realization of the importance of protein motion in biological processes, thanks also to the availability of methods that can capture some glimpses of the ever-changing cellular activity.
Molecular motion (here intended as relative motion of their constituent atoms, not as mere translocation of the protein as a whole object to different places in the cell) can be studied from several different perspectives: X-ray crystallography studies can provide static images of proteins in different conditions (for example with and without a ligand, or in association with other proteins); Nuclear Magnetic Resonance (NMR) capture varying conformations of proteins in solution; molecular dynamics simulations can provide a series of states, sometimes very different, in particular when 'directed' or associated with other information; indirect techniques like FRET can also give indication on relative positioning of protein domains, and their dynamics; electron microscopic techniques, X-ray microscopy and other techniques, all contribute to adding new and important information that can advance our interpretation of biology in action at the near atomic level.

NMR spectroscopy exploit the properties of nuclear resonance of protons or other excited nuclei to identify conformations of proteins in solution. This technique allows the capture of a number of conformations, not constrained by the strict alignment of a crystal, providing a glimpse on the flexibility and the motion allowed by the relative constraints derived by the primary and secondary structure.
Of the >60.000 protein structures in the Protein Data Bank [2, 3], about 13% were obtained by NMR spectroscopy. These typically contain a collection (10-50) of models in no specific order. It is logical to think that at least some of the models belong to a sequence, which, seen in a flow, could reveal the overall movement of the protein.
Based on these considerations, we have set up a procedure that allows the 'ordering' of models in a NMR set, and at the same time provides intermediate conformations that smoothly transit the protein between them. For this, we have developed BioBlender, a set of features designed to be used by biologists, based on Blender [4]. Blender is a complete package for 3D animation, Game Engine (GE), and rendering, developed as an Open Source project by the Blender Foundation and scriptable in Python. In Blender, the GE instances the Bullet physics library [5], an open source software multi threaded 3D collision detection, soft body and rigid body dynamics library (physics engine) for video games and visual effects. Blender GE uses a system of graphical 'logic bricks' (a combination of 'sensors', 'controllers' and 'actuators') to control the movement and display of objects in the engine. For our aims, the essential features of the engine are the collision detection, the control of rotation with the rigid body constraints, and the capability of baking (recording) the movements of objects as calculated during game playing.

## Results

As the test material for our procedure, we have used the 25 models of Calmodulin (CaM) stored in the pdb file 1cfc [6]. CaM is a well studied protein composed of 148 aminoacids, comprising 1166 atoms (2262 including Hydrogens), well conserved along the evolutionary scale [7]. CaM is composed of two globular domains (heads) connected by a flexible linker. Each domain is also mobile: it is composed of four alpha helices, organized in two EF hand-Calcium binding motifs, which undergo a major transition upon Ca binding, but are also quite flexible in solution in the absence of this ion.

The procedure to obtain sequential ordering and motion of the protein follows the steps shown in Figure 1.

First, using the script *spdbv_rmsd* we calculated all RMSD (all atoms) between all models in the NMR collection (Supplementary file Table S1). We selected the two most distant conformations (7 and 21, rmsd 11.87 Å), and uploaded them in BioBlender (all atoms; inclusion of Hydrogens significantly increases the calculation times, with marginal improvements on resulting structures) as two different positions of the same 'object', and separated the two by an arbitrary N of frames (100). Frames are considered discrete steps: the name derives from the cinema setting, in which, in order to obtain a smooth vision, it is necessary to display at least 24 frames per second. BioBlender (Figure 2) builds the molecule in the 3D environment by creating a sphere for each atom (with its covalent radius or with a collision radius equivalent to the specific Van der Waals radius), and links (corresponding to chemical bonds, built using an aminoacid library) which are set as rigid body joints, allowing rotation along their own axis.

With this settings, the Blender GE is played and the positions of all atoms at all intermediate frames are recorded. These are exported in a series of files in pdb format (command *export PDB* of BioBlender interface, Figure 2), and compared by rmsd with all the remaining models in the original NMR collection (script *spdbv_rmsd*). Data were plotted in graphic form (Figure 3), and any model found similar (RMSD $\geq$ 2 A) to one of the intermediates calculated by BioBlender, was considered a step in the route between the two distant models.

The procedure was then repeated, this time introducing in BioBlender the start and end conformations plus the ones found close to the path. By reiterating the process using PDB entry 1cfc, and models 7 and 21 as starting conformations, we found 'on the way' between them, models 19, 22, 13, 3 and 1. Thus we ordered about one third of the conformations (Figure 4). We started again, this time using two other models as starting points.

In this way, a map was built that allows all models of the NMR collection to be reached, starting from any other one. We have plotted the map in a 3D graph made with Jmol (Supplementary File 2), in which all conformations are linked, see Figure 5.

The question of the physical and chemical plausibility of the intermediate steps calculated by Blender GE is addressed: all the .pdb files of the sequence were exported and analysed with Swiss-PdbViewer (spdbv) [8, 9], as shown in Figure 6.
We developed a set of scripts (see below) to perform some functions automatically in spdbv, since this program cannot be accessed through BioBlender interface. Energy content of each conformation was evaluated by GROMOS 43B1 [10] force field included in spdbv (script *spdbv_energy*) and plotted. A close examination of the contribution to total energy revealed that most peaks were due to minor geometrical distortions, mainly due to rotameric conversions, or to close proximity of atoms. Rotamers were manually adjusted, by inverting the names of the equivalent atoms involved, and distorted geometries were fixed through energy minimization, performed within spdbv program using again the GROMOS force field. This step was automated using the script *spdbv_minimiz*, which allows users to set a range of energy around which to stop the minimization cycles. We set such value to be of the same order of those of the NMR models (Figure 7A). The script also outputs the number of minimization cycles necessary to obtain structures with such energy content. Cycles were typically less than 50 (Figure 7B); however, when this value exceeds 50, it is indicative of a specific problem. The mean displacement of atoms necessary to reach such conformations, starting from the Blender calculated positions, was typically less than 0.02 Å .
The entire sequence of adjusted files was finally reimported in BioBlender, (one conformation every frame), so that it could be used to visually inspect the moving protein in a 3D environment.
The sequence of files was also used to feed the visualization scheme described in the accompanying paper.

## Conclusions

In this report, we provide a contribution to the interpretation of data derived from NMR: namely, we demonstrate how using the power of 3D computer graphics techniques, in particular the game engine, it is possible to interpolate between many different models of a protein, stored in the NMR file, and obtain simultaneously both a series of intermediates and a navigation map that permits the direct inspection of the protein in its motion.
The procedure can be performed using mainly BioBlender interface, together with other necessary programs of general use in most structural biology laboratories. The user is prompted for information on file location, choice of parameters (N of Models to upload, N of frames between models, atoms to import etc.) and is updated on the work progress, or notified in case of errors.
The system is suited for the elaboration of transition states between models which are not very different, such as those obtained by NMR studies; however the same principle can be applied to larger motion, for example between structures obtained in different biochemical conditions.

### Programs and Scripts

Programs and Scripts written by us are provided as supplementary files

**BioBlender** – the engine described in this paper, based on Blender 2.5. It is provided as supplementary material, but we suggest to download the most updated version from the website www.BioBlender.net, where users can also find tutorials, and a forum for open discussion of further developments and (quite certainly) handling of errors.
**Blender 2.5** – a free, open source, cross platform suite of tools for 3D creation [4].
**Python 2.5** – an interpreted, interactive, object-oriented, extensible programming language [11].
**Swiss-PdbViewer** – General purpose Molecular Software [8, 9] that also allows to analyze several proteins at the same time, superimposing and comparing them.

*spdbv_rmsd* Script to calculate rmsd between different set of .pdb files using spdbv
*spdbv_energy* Script used in spbdv to automate batch analysis of pdb files
*spdbv_minimiz* Script used in spdbv to optimize the geometry of the intermediates calculated by Blender GE

## Availability and Future Directions

Project name: BioBlender
Project download page: www.scivis.ifc.cnr.it, www.bioblender.net
Operating system: Windows
Requirements: install PyMOL, python and numpy (they can be found in the Installer folder). These requirements apply to the complete BioBlender, including for the functions described in the accompanying paper [1].
We are currently developing more complete version of the software that will include:

- libraries for importing and working with nucleic acids and other molecules
- Linux compatibility (now it can be run on Linux using Wine)
- new algorithm to solve the rotamer problem, also suitable for major protein movements.

An interesting possibility would be to control all functions of Swiss-PdbViewer through the BioBlender interface, or to use an alternative instrument for the physics evaluation of BioBlender-calculated models.

## Acknowledgments

We thank Giuseppe Maraziti for help with programming, Stefano Cianchetta and Maria Antonietta Pascali for assistance in the generation of figures, the Swiss-PdbViewer team for help online and the Blender users and developers community for kind answers to our many questions.

## Figure Legends

**Fig. 1**
Scheme of procedure for generating sequence of NMR models
**Fig. 2**
BioBlender interface
**Fig. 3**
RMSD of Blender-calculated transition conformations (frames 1-100) with all Models in original NMR file
**Fig. 4**
RMSD of the sequence of models transiting the protein between conformations 7 and 21 of 1cfc, with all models in the 1cfc NMR file
**Fig. 5**
The navigation map of NMR file 1cfc. The image is a view of the map in Jmol (Supplementary File 2)
**Fig. 6**
Scheme of file processing for physico-chemical evaluation
**Fig. 7**
**A** Number of minimization cycles by GROMOS to obtain conformation with energy content of about -2500 KJ/mol
**B** Energy content for each intermediate frame of the sequence 7-19-22-13-3-21.

## Supplementary material

All material can be downloaded as a compressed folder from this link .

**Table S1**
all rmsd of 1cfc

**File S2**
Jmol file of navigation map. Folder containing: READ_ME.txt; Jmol.pov (open with Jmol, contains the NMR map); graph3D.txt text file (paste in Jmol console to obtain the Jmol map)

**File S3**
BioBlender. Please download from the link above (program cannot be uploaded as Supplementary File).

**File S4**
*Script spdbv_rmsd.* Script used in spdbv to perform automatic comparison of rmsd between several pdb files. Complete instructions in the README file.

*Script spdbv_energy*. Script used in spdbv to analyze .pdb files created by Blender game engine. Complete instructions in the README file.

*Script spdbv_minimiz*. This script is used from within the spdbv console.
You should enter few parameters and it will output a series of minimized files, and other information. Complete instructions in the README file.

# Figure 1

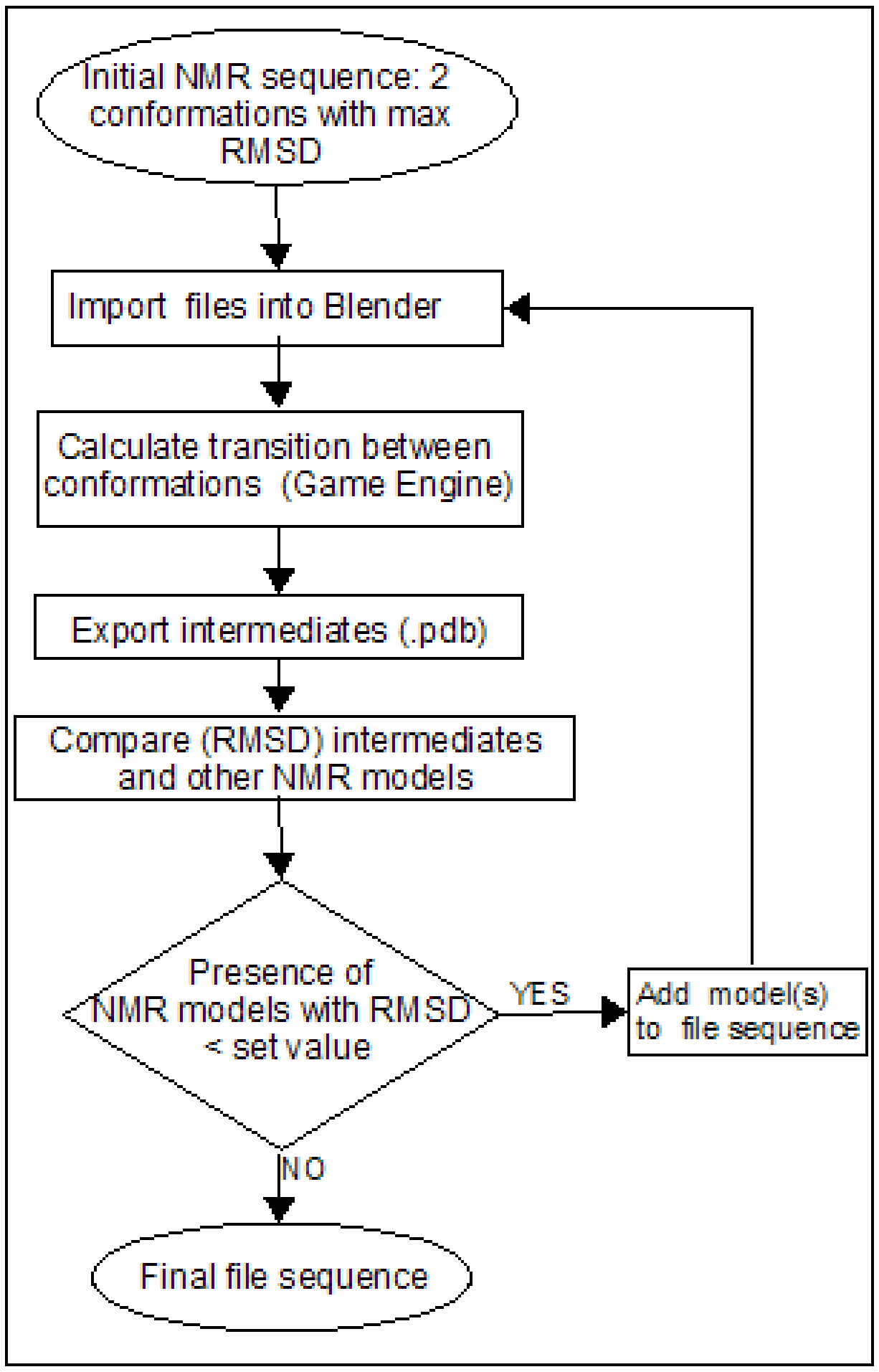
Initial NMR sequence: 2 conformations with max RMSD
Import files into Blender
Calculate transition between conformations (Game Engine)
Export intermediates (.pdb)
Compare (RMSD) intermediates and other NMR models
Presence of NMR models with RMSD < set value
YES
Add model(s) to file sequence
NO
Final file sequence

**Figure 2**

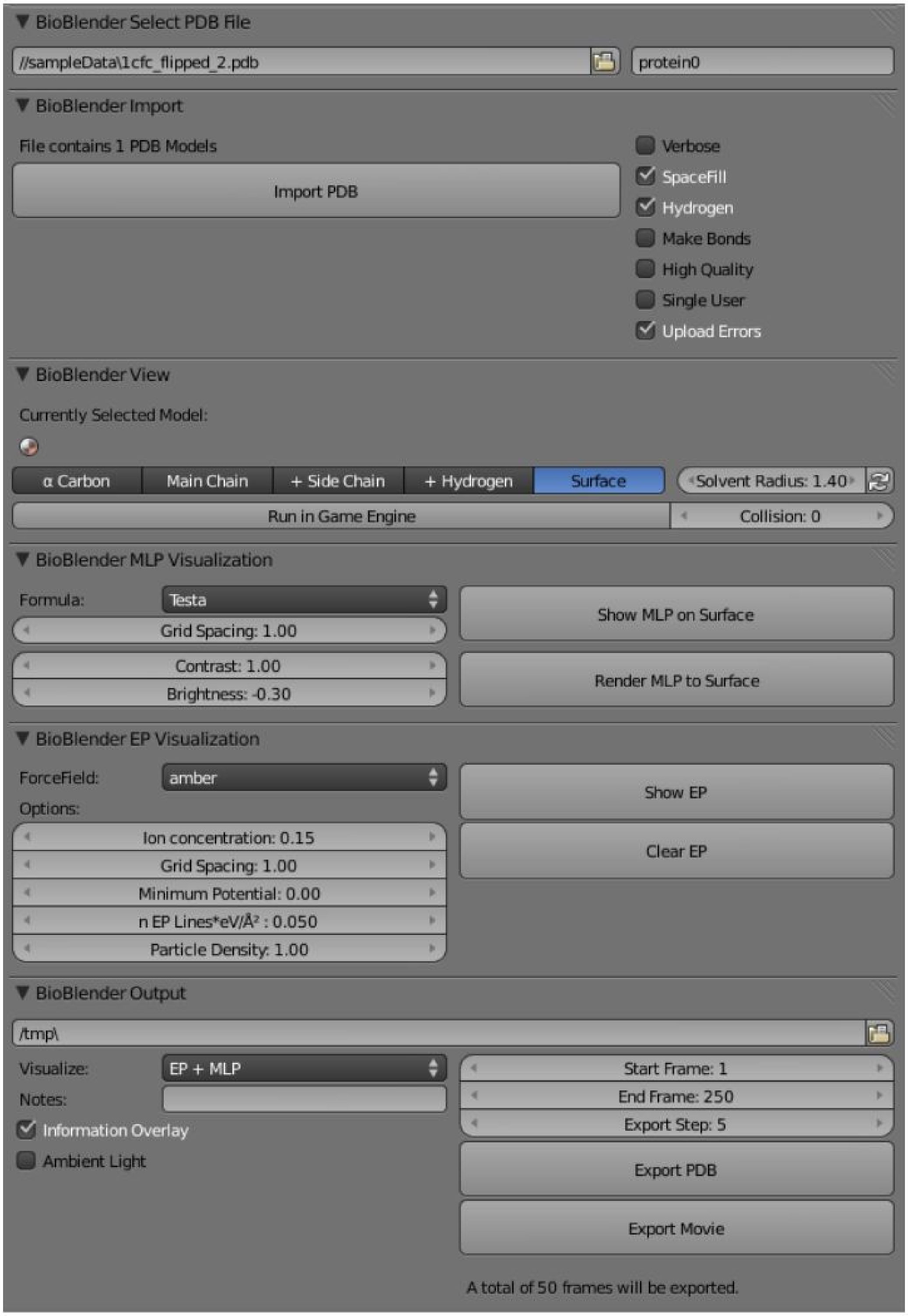
BioBlender Select PDB File
//sampleData\1cfc_flipped_2.pdb
protein0
BioBlender Import
File contains 1 PDB Models
Import PDB
Verbose
SpaceFill
Hydrogen
Make Bonds
High Quality
Single User
Upload Errors
BioBlender View
Currently Selected Model:
α Carbon
Main Chain
+ Side Chain
+ Hydrogen
Surface
Solvent Radius: 1.40
Run in Game Engine
Collision: 0
BioBlender MLP Visualization
Formula:
Testa
Grid Spacing: 1.00
Contrast: 1.00
Brightness: -0.30
Show MLP on Surface
Render MLP to Surface
BioBlender EP Visualization
ForceField:
amber
Options:
Ion concentration: 0.15
Grid Spacing: 1.00
Minimum Potential: 0.00
n EP Lines*eV/Å² : 0.050
Particle Density: 1.00
Show EP
Clear EP
BioBlender Output
/tmp\
Visualize:
EP + MLP
Notes:
Information Overlay
Ambient Light
Start Frame: 1
End Frame: 250
Export Step: 5
Export PDB
Export Movie
A total of 50 frames will be exported.

**Figure 3**

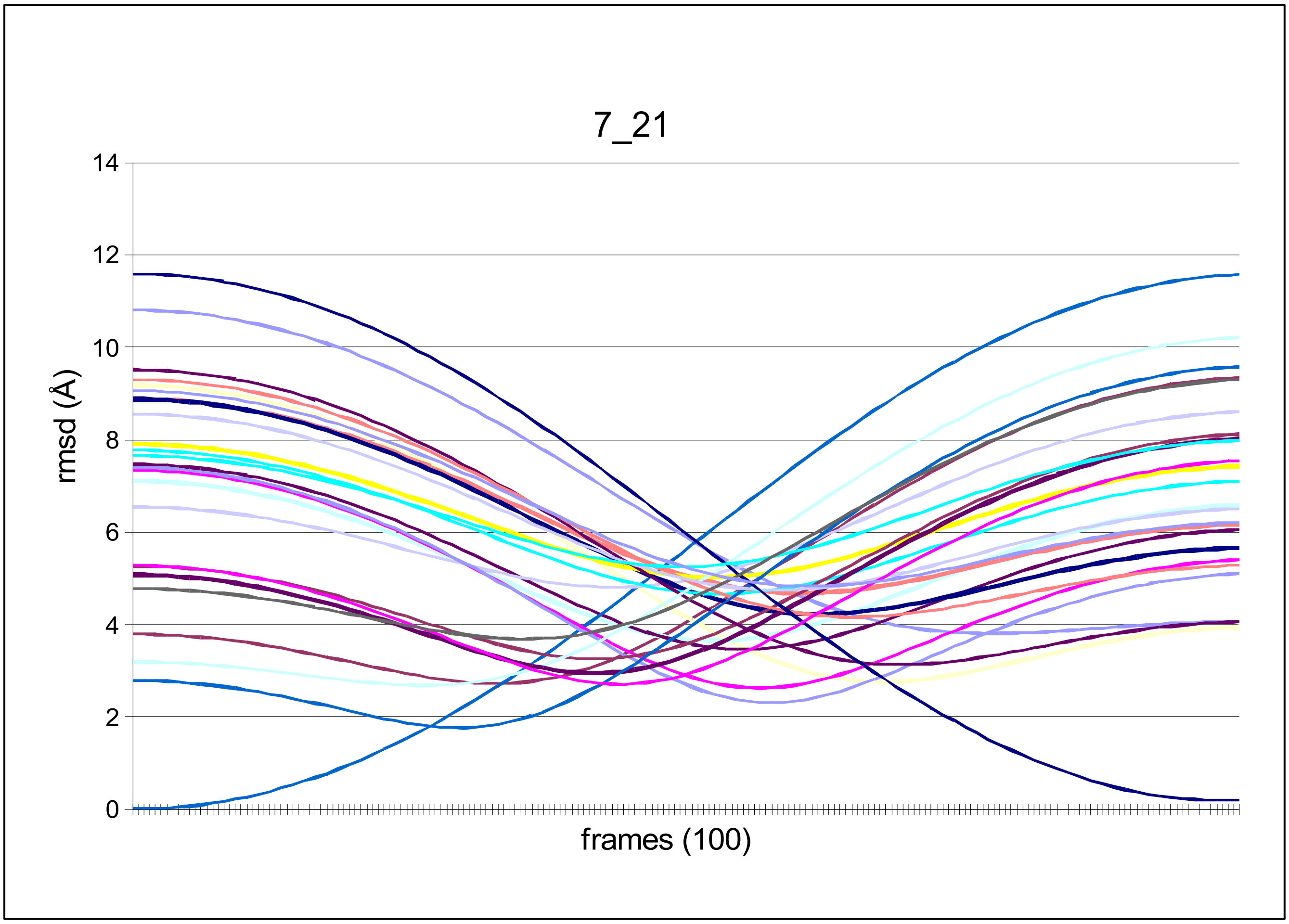

# Figure 4

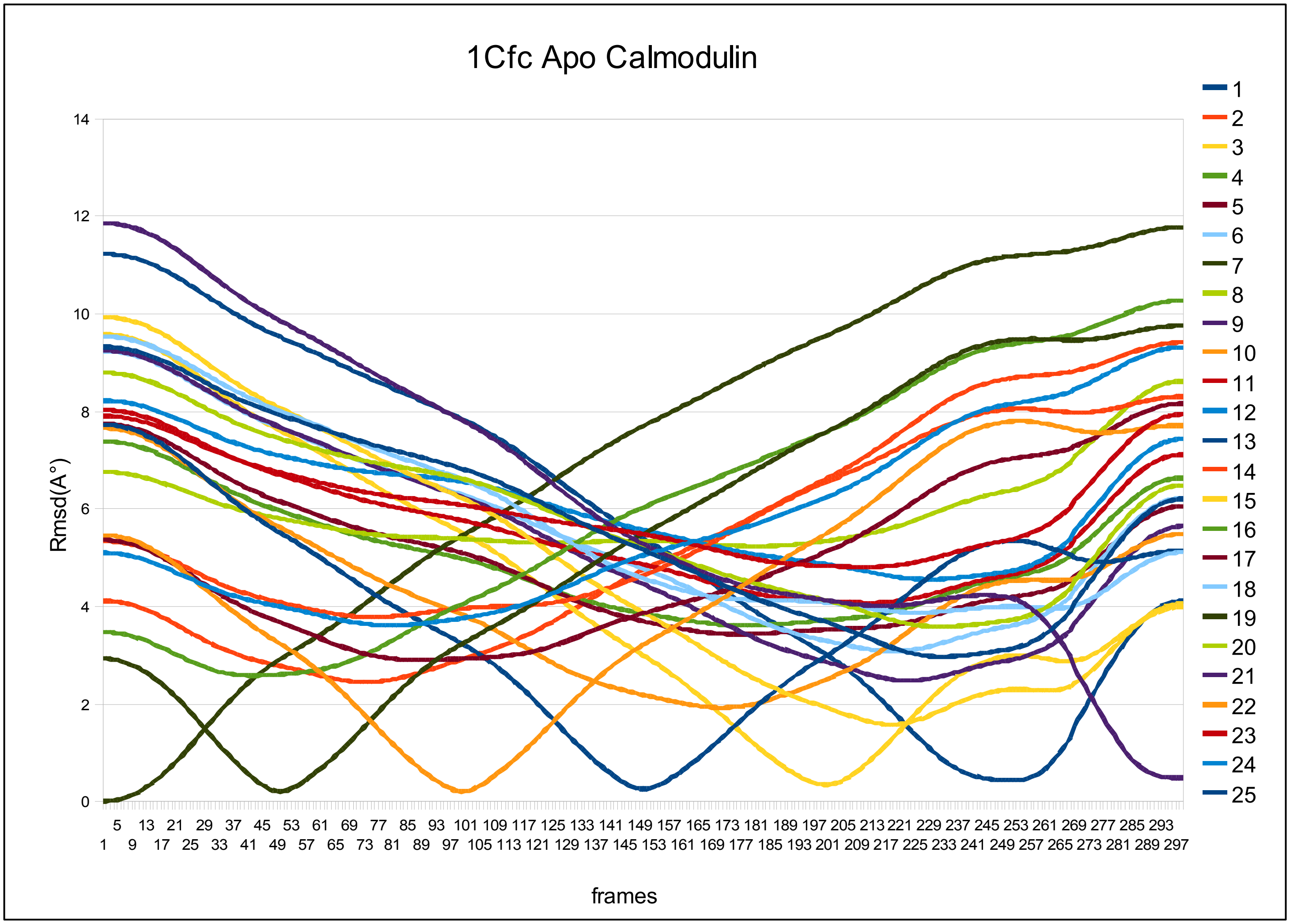

**Figure 5**

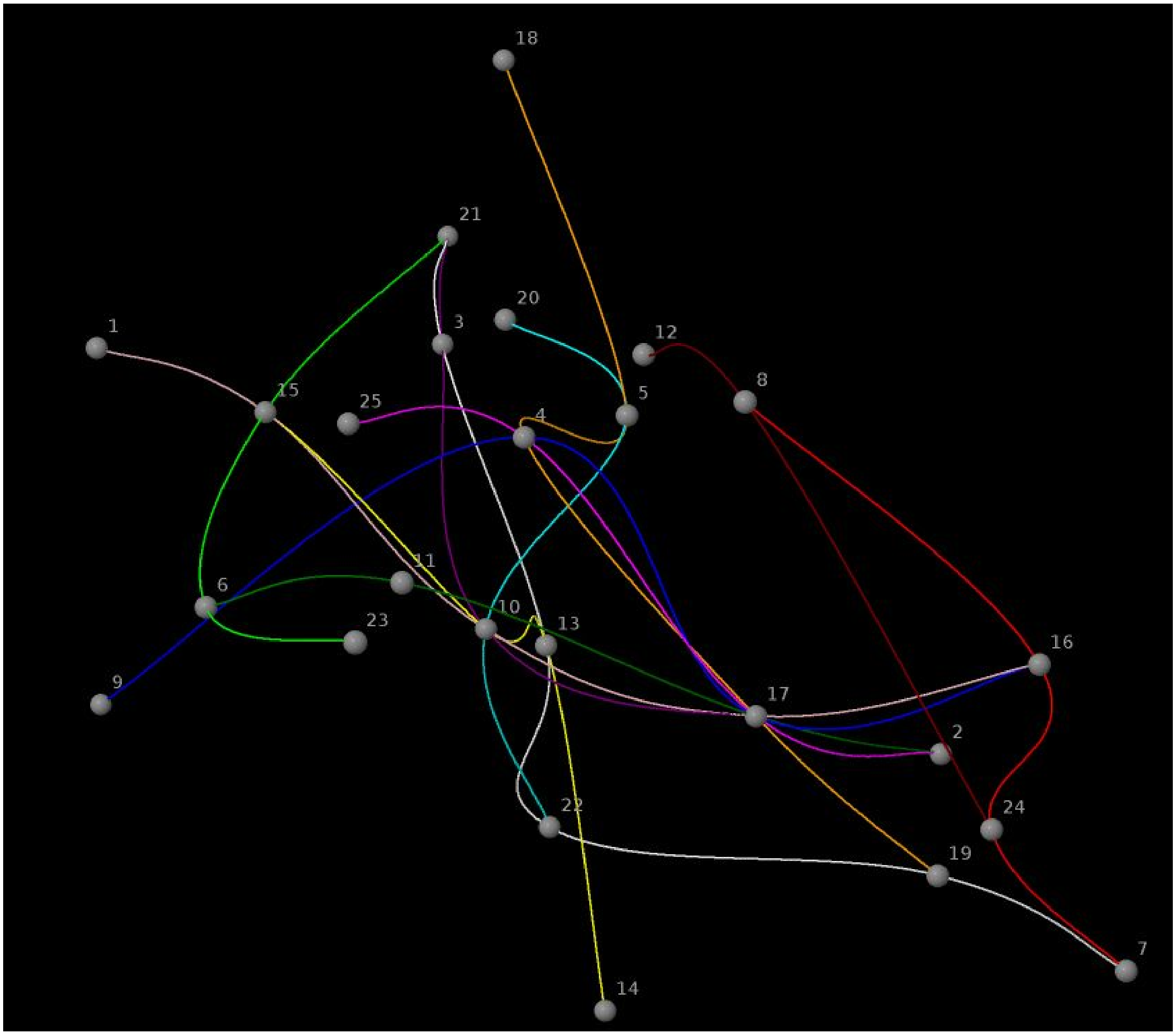

**Figure 6**

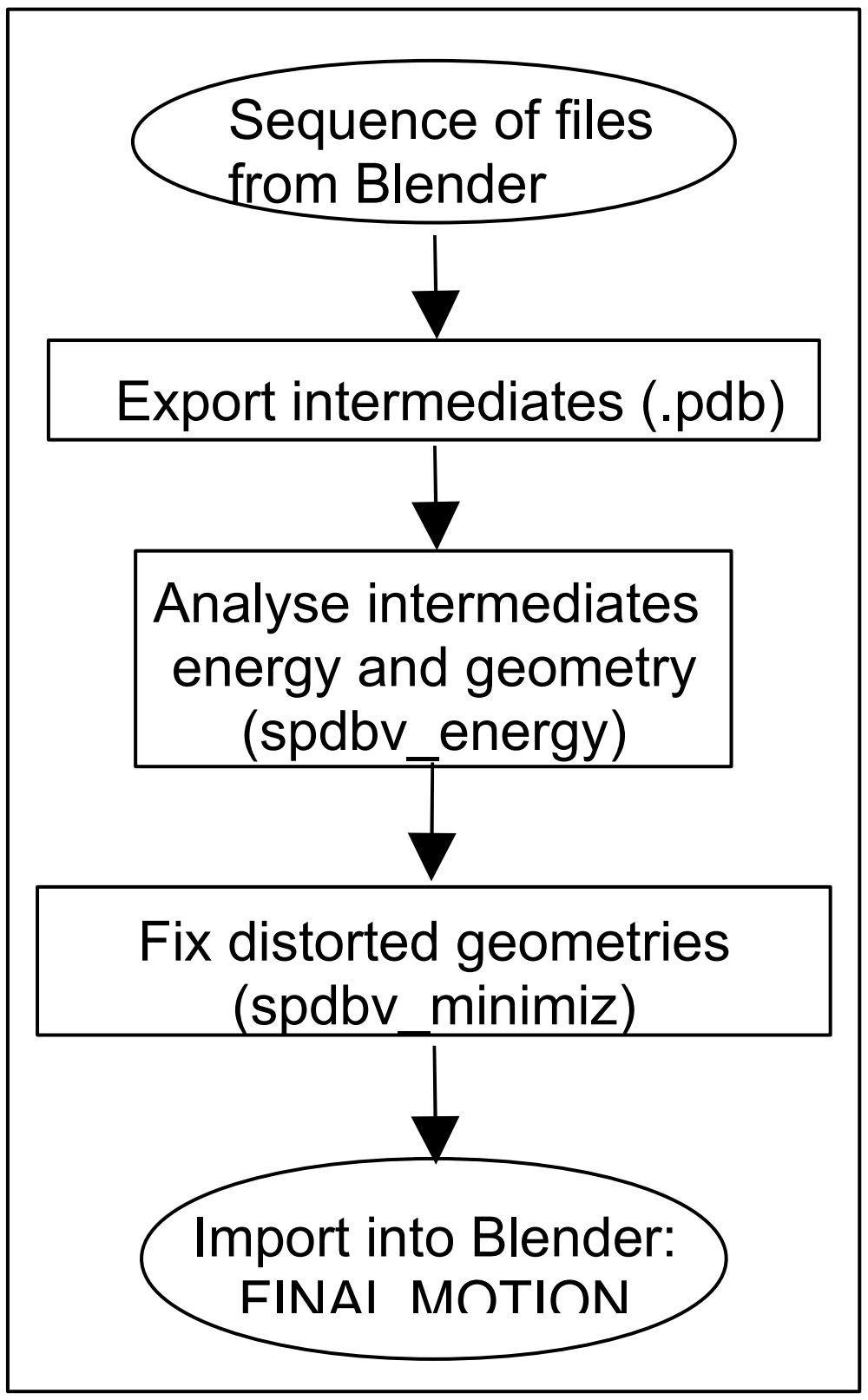

# Figure 7

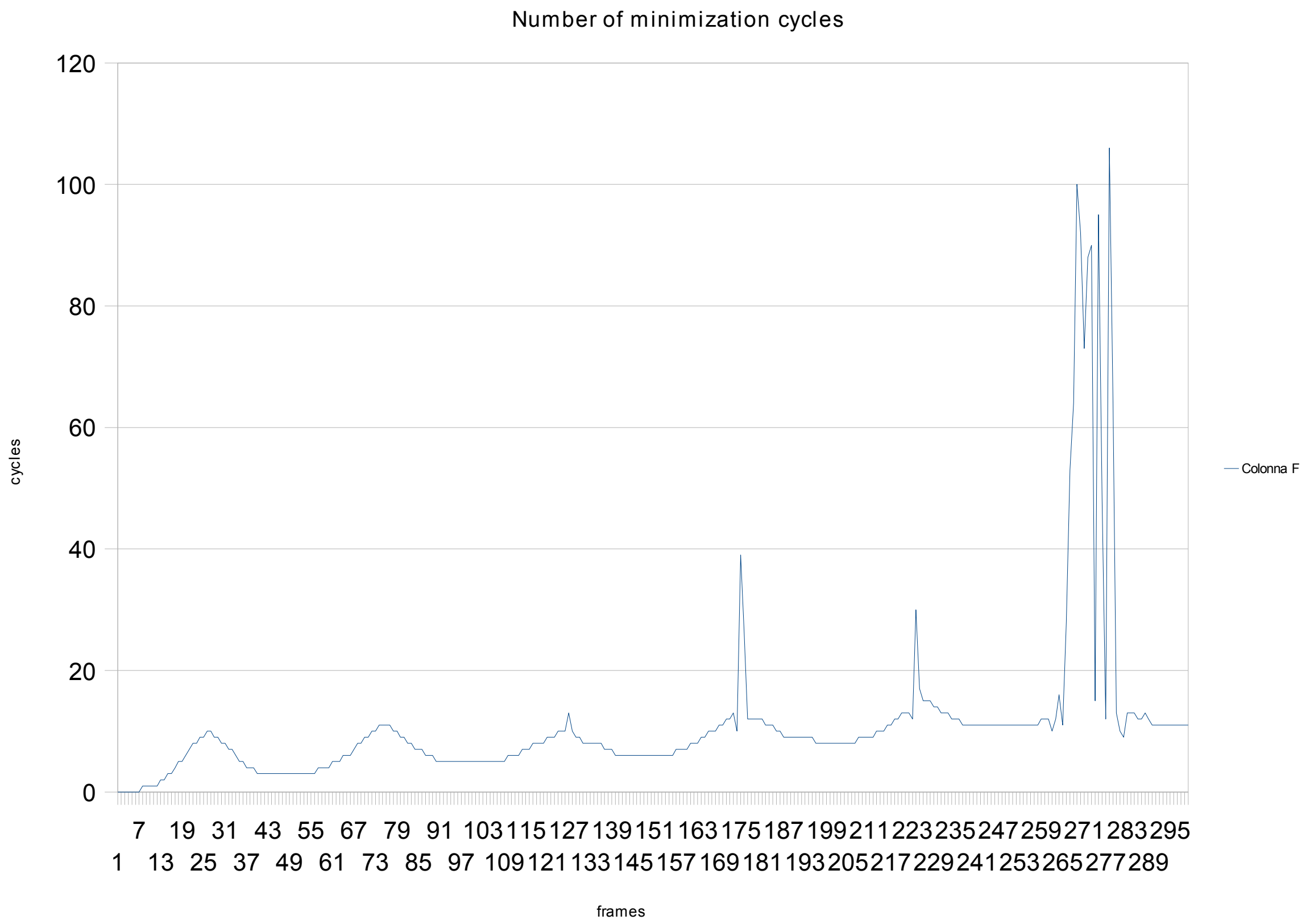
Number of minimization cycles
cycles
120
100
80
60
40
20
0
7 19 31 43 55 67 79 91 103 115 127 139 151 163 175 187 199 211 223 235 247 259 271 283 295
1 13 25 37 49 61 73 85 97 109 121 133 145 157 169 181 193 205 217 229 241 253 265 277 289
frames
Colonna F

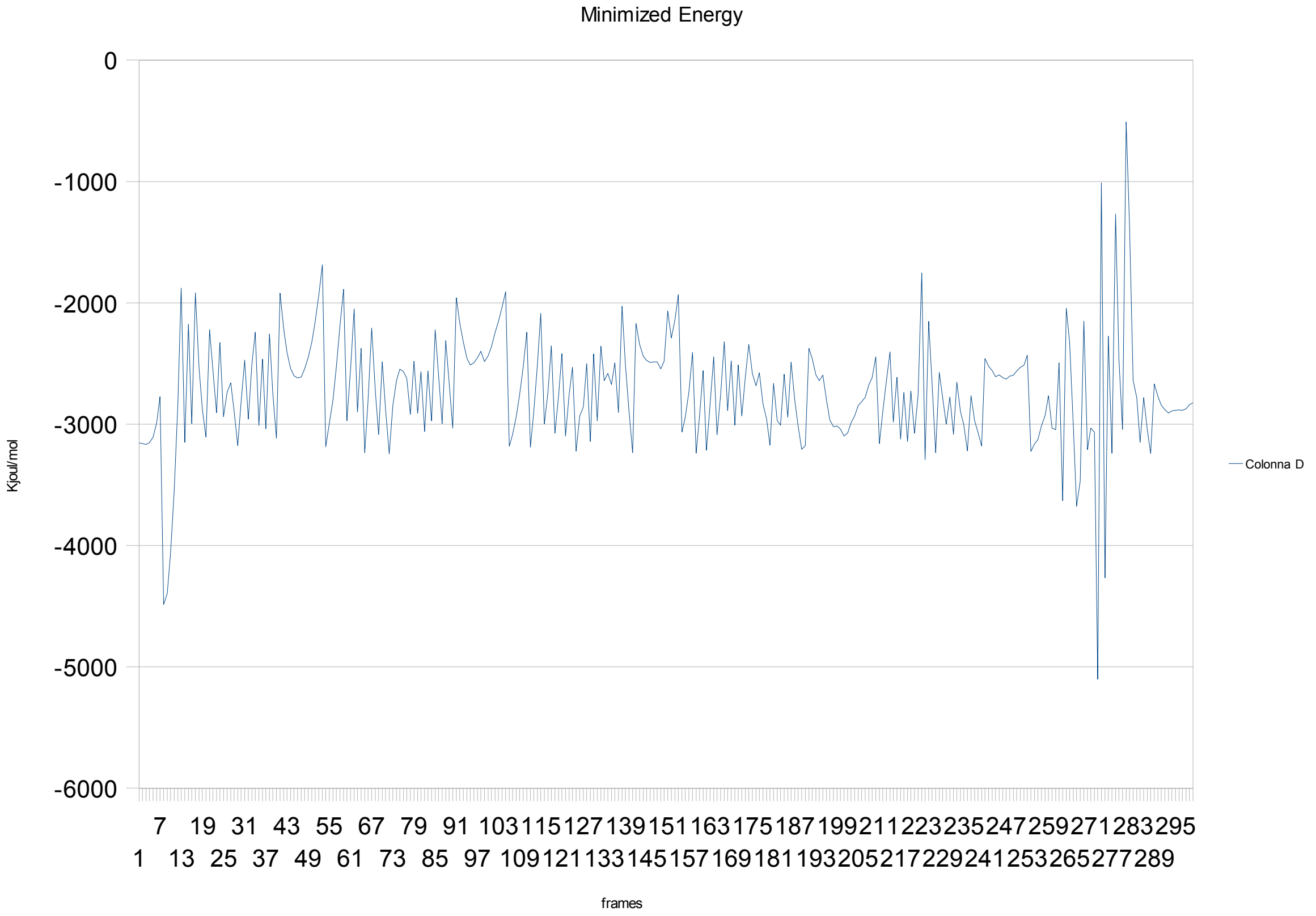
Minimized Energy
Kjoul/mol
0
-1000
-2000
-3000
-4000
-5000
-6000
7 19 31 43 55 67 79 91 103 115 127 139 151 163 175 187 199 211 223 235 247 259 271 283 295
1 13 25 37 49 61 73 85 97 109 121 133 145 157 169 181 193 205 217 229 241 253 265 277 289
frames
Colonna D